\documentclass[useAMS,usenatbib]{mn2e}
\usepackage{graphicx,color,epsfig}
\usepackage{amsmath, amssymb}
\usepackage{mathrsfs}
\usepackage{url}

\newcommand{\be}{\begin{equation}}
\newcommand{\ee}{\end{equation}}

\def\ltsima{$\; \buildrel < \over \sim \;$}
\def\simlt{\lower.5ex\hbox{\ltsima}}
\def\gtsima{$\; \buildrel > \over \sim \;$}
\def\simgt{\lower.5ex\hbox{\gtsima}}
\newcommand\sgra{Sgr~A$^*$}

\newcommand\ledd{{L}_{\rm Edd}}

\def\del#1{{}}

\def\msun{{\,{\rm M}_\odot}}
\newcommand\mbh{{\,{\rm M}_{\rm bh}}}

\def\lsun{{\,L_\odot}}

\title[\sgra\ flares: tidal disruption of asteroids and
  planets?]{\sgra\ flares:  tidal disruption of asteroids and planets?} 

\author[Kastytis Zubovas, Sergei Nayakshin and Sera Markoff]
       {\parbox{18cm}{Kastytis Zubovas$^{1,*}$, Sergei Nayakshin$^1$ \& Sera Markoff$^2$}\vspace{0.3cm}\\
\noindent $^1$ Dept. of Physics \& Astronomy, University of Leicester,
Leicester, LE1 7RH, UK\\
\noindent $^2$ Astronomical Institute 'Anton Pannekoek', University of
Amsterdam, P.O. Box 94249, 1090 GE Amsterdam, the Netherlands}

\begin{document}

\maketitle

\begin{abstract}
It is theoretically expected that a supermassive black hole (SMBH) in
the centre of a typical nearby galaxy disrupts a Solar-type star every
$\sim 10^5$ years, resulting in a bright flare lasting for months. Sgr
A$^*$, the resident SMBH of the Milky Way, produces (by comparison)
tiny flares that last only hours but occur daily. Here we explore the
possibility that these flares could be produced by disruption of
smaller bodies -- asteroids. We show that asteroids passing within an
AU of \sgra\ could be split into smaller fragments which then vaporise
by bodily friction with the tenuous quiescent gas accretion flow onto
\sgra.  The ensuing shocks and plasma instabilties may create a
transient population of very hot electrons invoked in several
currently popular models for \sgra\ flares, thus producing the
required spectra.  We estimate that asteroids larger than $\sim 10$ km
in size are needed to power the observed flares, with the maximum
possible luminosity of the order $10^{39}$ erg s$^{-1}$. Assuming that
the asteroid population per parent star in the central parsec of the
Milky Way is not too dissimilar from that around stars in the Solar
neighborhood, we estimate the asteroid disruption rates, and the
distribution of the expected luminosities, finding a reasonable
agreement with the observations. We also note that planets may be
tidally disrupted by \sgra\ as well, also very infrequently. We
speculate that one such disruption may explain the putative increase
in \sgra\ luminosity $\sim 300$ yr ago.

\end{abstract}

\renewcommand{\thefootnote}{\fnsymbol{footnote}}
\footnotetext[1]{E-mail: {\tt kastytis.zubovas@astro.le.ac.uk }}

\section{Introduction}

Most of the nearby SMBHs are rather dim \citep[e.g.,][]{Ho2008ARA&A},
suggesting that little gas is supplied to them at the current
epoch. However, rare and temporary exceptions from this ``gas
drought'' are expected to occur when a star passing too close to a
SMBH is shredded into streams by the tidal forces of the SMBH
\citep{Rees1988Natur}. The bound streams precess and self-intersect on
the return passage past the black hole, resulting in very strong
shocks. The result of these shocks should be a small-scale accretion
disc around the SMBH, and thus power a spectacularly bright $L\sim
10^{44}-10^{46}$ erg s$^{-1}$ flare, lasting a few months
\citep{Lodato2009MNRAS}. Such candidate events have indeed been
obsered in nearby galaxies \citep{Esquej2008A&A}, and there is one
recent observation of a $\gamma$-ray source that is best explained by
a jet resulting from a stellar tidal disruption event
\citep{Bloom2011arXiv}.

Tidal disruption events are unlikely to be observed from \sgra\ any time soon
because they are expected to be quite rare, i.e. $\dot{N}_{\rm tid} \sim
10^{-5}$~yr$^{-1}$ per galaxy \citep{Alexander2005PhR}. However, the centre of
our Galaxy does produce enigmatic flares on a roughly daily basis. Although
miniscule in amplitude ($L_f \sim 10^{34}-10^{35}$ erg s$^{-1}$) when compared
with stellar disruptions, \sgra\ flares are still $10-100$ times brighter than
its quiescent state. The flares are also shorter, lasting hours rather than
months. Here we explore a new scenario for the flares: that they are the
result of tidal disruption of asteroids rather than stars. The fact that there
are many more asteroids than stars and that the asteroids are much smaller
than stars would naturally explain why \sgra flares are much more frequent but
much less luminous and shorter than the stellar tidal disruption events.

In this paper we test the asteroid disruption hypothesis for
\sgra\ flares in a reasonable level of detail. In doing so, we adopt
an approach complimentary to most of the exisiting popular models of
\sgra\ flares. As reviewed in \S \ref{sec:model} below, these usually
predict spectra given specific assumptions about emitting particle
distributions; it is not always specified how these distributions are
energized. In the context of our model, instead, there is far too much
physical uncertainty in predicting the particle distributions at this
stage, but we are able to constrain the energetics, the duration and
the frequency distribution of the tidal disruption events starting
from reasonable assumptions about the populations of asteroids
in the central parsec of \sgra. Our model presents a mechanism
  for producing the transient hot particle populations responsible for
  the observed flares.

The paper is structured as follows. In Section \ref{sec:model} we
overview the astrophysical setting of the problem, and the
observational characteristics of the flares. In Section
\ref{sec:destr}, we estimate the minimum size of the asteroids ($\sim
10$ km) needed to power the observed flares.  We then consider what
happens to asteroids of different sizes as they pass by \sgra\ on
orbits of a given pericentre distance. We show that large asteroids
approaching \sgra\ within $R\simlt 1$ AU are broken into smaller
pieces (at most $\sim1$ km in size). We also point out that asteroids
evaporate as they pass through the gas of the tenuous quasi-spherical
accretion flow \citep[that is believed to power the quiescent \sgra
  emission][]{Narayan1995Natur, Yuan2003ApJ} at very high
velocities. The combination of tidal ``grinding'' of large asteroids
into smaller fragments and evaporation of the latter may destroy the
asteroids efficiently and turn their bulk energy into heat in the
shocks between the evaporated material and the background accretion
flow.

In Section \ref{sec:supply} we calculate the rate at which asteroids are
supplied into the vicinity of the SMBH and find values roughly consistent with
the frequency of observed flares. In Section \ref{sec:planets} we note that
planets, too, could be tidally disrupted by \sgra, although clearly far less
frequently than asteroids. We consider whether one such disruption could
account for the suspected \sgra\ brightenning to $\sim 10^{39}$ erg s$^{-1}$
$\sim 300$~yr ago due to the well-known X-ray echo on Sgr B2 molecular cloud
\citep{Sunyaev1998MNRAS,Revnivtsev2004A&A}, that is now fading
\citep{Terrier2010ApJ}.  Finally, in \S \ref{sec:flare}, we suggest how the
evaporating asteroids could produce high energy particles needed by the
current models of flare emission from \sgra. We present a summary discussion
and conclusions of the model in Section \ref{sec:discuss}.

\section{\sgra\ and its flares} \label{sec:model}

Sgr A$^*$ is the supermassive black hole (SMBH) in the nucleus of our
Galaxy, with the mass $\mbh \simeq 4 \times 10^6 \msun$
\citep{Schodel2002Natur, Ghez2005ApJ}. By comparison with active
galactic nuclei (AGN) \sgra\ is famously dim in all frequencies. Its
bolometric luminosity is only $L_{\rm bol} \simeq 300 \lsun \simeq
10^{-9} L_{\rm Edd}$ \citep[e.g.,][]{Melia2001ARA&A}.  In X-rays 
  \sgra \'s quiescent luminosity is less than $\sim 10^{-11} \ledd$,
where $\ledd \sim$ a few $\times 10^{44}$ erg s$^{-1}$ is its
Eddington luminosity \citep{Baganoff2003ApJ}, and in the near infrared
$L \sim 10^{35}$ erg s$^{-1}$ \citep{Genzel2003Natur}. This
extraordinarily low luminosity has been explained in the literature via
models of radiatively inefficient inflow and/or outflow \citep[and
  references therein]{Narayan1995Natur, Falcke2000A&A,
  Narayan2002luml.conf, Yuan2003ApJ}.

The quiescent emission from \sgra\ is punctuated several times a day
by short flares in the near infrared \citep{Genzel2003Natur,
  Ghez2004ApJ, Marrone2008ApJ}. Approximately once per day, these
flares are accompanied by corresponding rises in the X-ray emission
\citep{Baganoff2001Natur, Eckart2006A&A, Hornstein2007ApJ,
  Marrone2008ApJ, Porquet2003A&A, Porquet2008A&A}. When both NIR and
X-ray flares occur, they are almost certainly causally connected and
show no appreciable time lag between their peaks, although IR
lightcurves have shallower rising and decaying slopes
\citep{Hornstein2007ApJ, Yusef-Zadeh2006ApJ, Eckart2006JPhCS}. Sub-mm
flares have been observed approximately 1 hour later following some of
the IR/X-ray flares \citep{Mauerhan2005ApJ, Herrnstein2004AJ,
  Kunneriath2010A&A}, although the connection between the two has not
been firmly established.

Typically, flares last for approximately an hour to a few hours
($t_{\rm f} \lesssim 10^4$~s) and have luminosities a factor
$3$--$100$ above the quiescent emission level in both X-rays and near
infrared \citep[$L_{\rm X, NIR} \sim 10^{34} - 10^{35}$~erg
  s$^{-1}$;][]{Baganoff2001Natur, Genzel2003Natur}. There may also be
more frequent weaker flares that get blended in the quiescent emission
of \sgra \citep{Dodds-Eden2011ApJ}. The brightest flare observed so
far reached $L_{\rm f,max} \sim 10^{36}$~erg s$^{-1}$ in both the NIR
and the X-rays \citep{Porquet2003A&A, Porquet2008A&A}. The observed
NIR flare luminosity distribution (\S \ref{sec:size}) seems to follow
a $LN_{\rm L} \propto L^\alpha$ law, with $-1 \lesssim \alpha \lesssim
0$ \citep{Dodds-Eden2011ApJ}, where $N_{\rm L} \Delta L$ is defined as
the number of flares with maximum luminosity during the flare between
$L$ and $L +\Delta L$. The rise and fall times, as well as short
timescale variability, suggest that the flaring region is very compact
and located within $R \sim 10 R_{\rm S}$ of \sgra\, where $R_{\rm S} =
2GM_{\rm BH}/c^2 \sim 1.2\cdot10^{12}$~cm is the Schwarzschild radius
of \sgra \citep{Baganoff2001Natur, Porquet2003A&A, Shen2005Natur,
  Eckart2006JPhCS}. Besides the time variability constraints, the
location of the emission region is constrained directly by the NIR
observations to be within a few milliarcseconds of \sgra \citep[which
  is equivalent to tens of AU or a few hundred $R_{\rm
    S}$][]{Genzel2003Natur}.

There is currently no universally accepted model for \sgra\ flares.  Even the
emission mechanism is not completely settled. The suggested models are
synchrotron emission by either thermal or power-law distribution of electrons
for the NIR flares plus the inverse Compton or self-Compton emission in
the X-rays, or power-law synchrotron emission for all the components
\citep[e.g.,][]{Markoff2001A&A, Dodds-Eden2009ApJ}. In terms of associated
physical mechanisms responsible for flares, magnetic reconnection events
\citep{Yuan2003ApJ, Dodds-Eden2010ApJ}, turbulent shocks \citep{Liu2004ApJ}
and jet acceleration \citep{Markoff2001A&A, Yuan2002A&A, Maitra2009A&A} were
proposed. Short-timescale magnetic reconnection event models seem to be more
promising than transient density variation models
\citep{Markoff2001A&A, Dodds-Eden2010ApJ}.

Another class of flare models envisages a transient feature in the
accretion flow around \sgra. Such a feature may be an accretion
instability \citep[e.g.,][]{Tagger2006ApJ} or an orbiting hot spot
\citep[e.g.,][]{Broderick2005MNRAS}. Finally, a number of authors have
propsed an expanding plasma blob as the source of the flares
\citep{vanderLaan1966Natur, Yusef-Zadeh2006ApJb, Eckart2006A&A,
  Trap2011A&A, Kusunose2011ApJ}. A blob of relativistic plasma,
threaded by a magnetic field, is assumed to be suddenly created in the
accretion flow around \sgra and then proceeds to move outwards while
simultaneously expanding at a prescribed velocity. This leads to an
evolution of the optical depth of the plasma, which in turn causes
different parts of the emission spectrum to appear different during
the flare, leading to time lags between emission maxima and
characteristic light curves for the various spectral bands. The
orbiting hot spot model is similar to this, except that in the latter,
the plasma blob is assumed to circle around \sgra for at least several
dynamical times.

\citet{Nayakshin2004A&A} suggested that stars orbiting \sgra\ strike
an optically thick disc, and that the resulting shocks produce the
observed X-ray flares. This model is now firmly disfavoured by the
constraints on the NIR flaring region size of $\simlt 10$ AU, as it
would require unphysically large stellar densities in the innermost
region. Furthermore, no stellar eclipses or transient brightennings,
which would be two observable signatures of the optically thick disc
presence near \sgra\ \citep{Nayakshin2003MNRAS, Cuadra2003A&A} were
found either.

While stars cannot produce enough flares in the small region near \sgra, one
may legitimately wonder if disruption of asteroids instead of stars could
work\footnote{Scott Tremaine noted this point to one of us in about 2004}, as
there are far more asteroids than stars. We shall now turn to putting physical
constraints on this idea and argue that such a model may serve as a physical
basis for the 'expanding transient plasma blob' model (see above).

\section{Asteroid destruction near \sgra} \label{sec:destr}

\subsection{The minimum asteroid size}\label{sec:minimum}

We shall now estimate the minimum size of an asteroid necessary to
produce an observable flare. For the low end of observed flare
luminosities, i.e. $L_{\rm f} \sim 10^{34}$ erg s$^{-1}$ in both
X-rays and NIR, and the typical duration $\sim 10^4$ s. The resulting
total luminous energy release is $E_{\rm f} \sim 10^{38}$ erg in each
of these bands. This is the minimum energy that the asteroid should
produce upon interacting with the background gas flow around \sgra. We
assume that the energy is released by an asteroid of mass $M_{\rm a} =
4\pi/3 \rho_a r^3$, where $r$ is the asteroid's mean radius and
$\rho_{\rm a} \sim 1$ g cm$^{-3}$ is its material density \citep[see,
  e.g., Table 1 in ][]{Britt2002aste.conf}. It seems reasonable to
assume that the energy released in the flare is of the order of the
bulk energy of the asteroid.  Given that tidal disruption occurs
inside an AU of \sgra or so (see below), the bulk energy is a fraction
of the asteroid's rest mass. Thus, our estimate of energy released is
\begin{equation} 
\label{eq:eflare} E_{\rm f} = \xi M_{\rm a} c^2 = \frac{4 \pi
  \xi}{3} \rho_{\rm a} r^3 c^2 \sim 4 \cdot 10^{38} \xi_1 r_1^3
\; \mathrm{erg},  
\end{equation}
where $\xi = 0.1\xi_1$ is the dimensionless fraction of the asteroid's
rest mass energy released in the flare. We see that an asteroid with
$r \gtrsim 6$ km releases enough energy to power an observable flare,
if the whole energy is released in either IR or X-rays. For the rest
of the paper, we use a more conservative value $r \gtrsim 10$ km and
parametrise an asteroid's radius as $r \equiv 10 \; r_1$ km.

The brightest observed \sgra flare requires about a factor of 100 more
energy, which in our rough estimate would require an asteroid of $r
\sim 30$ km, with a more conservative estimate of $r \sim 45$ km.
Asteroids of these sizes are ``typical'' in the Asteroid belt of the
Solar System \citep{2005Icar..175..111B} and believed to lurk in the
extra solar debris discs as well \citep{Wyatt2008ARA&A}.

\subsection{Tidal disruption of an asteroid}\label{sec:td}

We shall consider large asteroids to have a ``rubber-pile'' structure,
i.e., be a collection of smaller rocks held together by gravity rather
than by material strength. This point of view is physically motivated
by the fact that large monolitic bodies are expected to collide at
high speeds with abundant smaller bodies. Such collisions do not
completely obliterate the large bodies but do erode them even in our
Solar System \citep{Chapman1978NASCP, Richardson1998Icar,
  Korycansky2006Icar}. In the environment we are considering,
collisions occur at even higher speeds (\S \ref{sec:collisions}), and
therefore the rubber-pile structure is even more relevant.

There are both similarities and differences in the way that asteroids
and stars are tidally disrupted near a SMBH. Since the mean density of
asteroids, $\rho_a$, is of the same order as that of main-sequence
solar type stars, the tidal disruption radius is very similar for
asteroids and stars. An asteroid is tidally disrupted in the vicinity
of the SMBH provided that
\begin{equation} 
\rho_{\rm a} \lesssim \frac{3 M_{\rm BH}}{4 \pi R^3}, 
\end{equation}
where $R$ is the distance to the SMBH. For $\rho_a = 1$ g cm$^{-3}$, the
tidal disruption radius is
\begin{equation}
R_{\rm td} \simeq 1.5 \cdot 10^{13} \; \mathrm{cm} \simeq 1 \; \mathrm{AU}.
\label{r_td}
\end{equation}

Unlike a star, a tidally disrupted asteroid breaks up into smaller
fragments that are bound by chemical forces rather than gravity.  The
fragments of the comet Shoemaker-Levy 9 tidally disrupted as it passed
by Jupiter are estimated to be around $\sim 1$ km in size \citep[see
  the discussion and references in][]{Asphaug1996Icar}. Through
analytical arguments and numerical simulations, \citet{Benz1999Icar}
suggest that objects greater than about $\sim 1$ km in diameter must
be composed of smaller pieces held together by gravity. We shall thus
consider the maximum size of the fragments to be around 1 km, and
probably less than 100 m due to a more extreme environment we study.

One further difference between stellar tidal disruptions and that of
asteroids is in the orbits of the disrupted
material. \citet{Rees1988Natur} shows that roughly half of the star's
material falls onto orbits bound to the black hole, whereas the other
half is ejected into the larger (outside $\sim 1$ pc) host galaxy. The
semimajor axes of the orbits of disrupted streams of gas can be found
from the specific energy of the relevant streams. Before the
disruption, the orbit is assumed to be parabolic, thus the specific
energy is nearly zero. After the disruption at pericentre distance
$R$, the specific energy of the stream is $\sim \pm v_{\rm a} \Delta
v_{\rm a}$, where $v_{\rm a}\sim 10^{10}$ cm/s is the parabolic
velocity of the object at the pericentre (eq. \ref{eq:va}), and
$\Delta v_a$ is the escape velocity from the object. The semimajor
axis of the most bound material is thus
\begin{equation} \label{eq:semi-axis}
R_{\rm orb} \sim \frac{G\mbh}{2 v_{\rm a} \Delta v_{\rm a}} \sim R
\frac{v_{\rm a}}{\Delta v_a}\;.
\end{equation}
For a Solar-type star, $|\Delta v_a|\sim$ few$\times 10^7$ cm
s$^{-1}$, and hence the semimajor axis of the most bound orbit is a
few hundred times the pericentre passage distance. This implies that
the material will fall back to the SMBH vicinity within a month to a
year, depending on the SMBH mass. This leads to a bright stellar
disruption flare \citep[for recent numerical simulations of the
  process see][]{Lodato2009MNRAS}.

However, for an asteroid, $\Delta v_a \sim 10^2 \; r_{\rm km}$ cm
s$^{-1}$, i.e., much smaller than for a star.  Therefore, if an
asteroid tidal disruption proceeded in exactly the same fashion as
that of a star, the change in the orbital energy of the different
fragments of the asteroid would be negligible. The disintegrated
asteroid would thus continue to travel on almost the same orbit as the
one it had before the disruption. The fragments would come back to the
SMBH after hundreds or thousands of years. As luminosity is energy
released per unit time, the luminosity output of such a disruption
would be far too small for us to be interested in it. Finally, unlike
the disrupted stellar gas streams, that are certain to intersect due
to precession of the orbits \citep{Rees1988Natur}, the returning
asteroid fragments are very unlikely to collide with one another. It
seems extremely unlikely that any significant flare would be produced
in this "dry" disruption scenario.

\subsection{Asteroid evaporation}\label{sec:vapor}

The inner few AU of our Galactic centre, or any other galactic centre,
are very likely to be filled with a gaseous accretion flow onto the
SMBH, however tenuous that flow might be. The asteroid moves through
this gas at almost a relativistic velocity. Aerodynamic friction may
cause a significant heating of the asteroid, perhaps leading to its
evaporation before it leaves the central region. We shall term this
background gas-mediated disruption ``wet disruption'' in contrast to
the dry disruption discussed in \S \ref{sec:td}.

The quiescent luminosity of \sgra\ and its linear polarization
measurements suggest an accretion rate $\dot{M} \gtrsim 10^{-8} \;
\msun$ yr$^{-1}$ in the system \citep{Aitken2000ApJ, Bower2003ApJ,
  Marrone2006ApJ}. If we assume that the flow is spherically
symmetric, and is in a free-fall onto \sgra, the gas density can be
estimated as
\begin{equation}
\begin{split}
\rho_{\rm g} &\simeq \frac{\dot{M}}{4 \pi R^2 v_{\rm ff}} = \frac{\dot{M}}{4
  \pi \left( G M_{\rm BH} R^3 \right)^{1/2}} \\ &\sim 3.4 \cdot 10^{-20}
\dot{M}_8 \; R_{\rm AU}^{-3/2} \; \mathrm{g \; cm}^{-3},
\end{split}
\end{equation}
where $\dot{M}_8 \equiv \dot{M}/10^{-8} \msun$ yr$^{-1}$. This is a
lower limit since a geometrically thick disc is a more plausible flow
configuration due to a likely nonzero angular momentum. Disc flows are
centrifugally supported and hence the radial velocity is always slower
than the free-fall velocity assumed above \citep{Shakura1973A&A,
  Narayan1994ApJ}. The results of \citet{Yuan2003ApJ} suggest a
density profile
\begin{equation}
 \label{eq:rhog}
\rho_{\rm g} \simeq \rho_0 \left(\frac{R}{R_{\rm S}}\right)^{-s} \; \mathrm{g}
\; \mathrm{cm}^{-3},
\end{equation}
with $\rho_0 \simeq 6.4 \cdot 10^{-17}$ g cm$^{-3}$ and $s \simeq
1.23$. Numerically,
\begin{equation}
 \label{eq:rhogAU}
\rho_{\rm g} \simeq 2.9 \cdot 10^{-18} \; \rho_{18} \; R_{\rm AU}^{-s} \;
\mathrm{g} \; \mathrm{cm}^{-3},
\end{equation}
where $\rho_{18}$ is a factor, of order unity, encompassing the possible
deviations from this model.  For definitiveness, we use equation
(\ref{eq:rhogAU}) in the calculations below.

An asteroid on a parabolic orbit close to the SMBH moves with velocity
\begin{equation}
 \label{eq:va}
v_{\rm a} \simeq \sqrt{\frac{2 G M_{\rm BH}}{R}} \simeq 9.4 \cdot 10^9 R_{\rm
 AU}^{-1/2} \; \mathrm{cm \; s}^{-1}.
\end{equation}
In the asteroid's rest frame, the mechanical energy flux of the background
accretion flow material striking the asteroid's surface is
\begin{equation}
\Phi_{\rm a} \sim \rho_{\rm g} v_{\rm a}^3 \simeq 2.4 \cdot 10^{12} \rho_{18}
\; R_{\rm AU}^{-3/2-s} \; \mathrm{erg} \; \mathrm{s}^{-1} \; \mathrm{cm}^{-2}\;.
\end{equation}
Assuming that a sizeable fraction of this energy flux is
reradiated as a thermal blackbody radiation, hence the effective temperature
of the asteroid is
\begin{equation}
T_{\rm a} \simeq \left(\frac{\Phi_{\rm a}}{\sigma_{\rm SB}}\right)^{1/4} \sim
1.4 \cdot 10^4 \rho_{18}^{1/4} \; R_{\rm AU}^{-3/8-s/4} \; \mathrm{K},
\end{equation}
where $\sigma_{\rm SB}$ is the Stefan-Boltzmann constant.  The
  radiation itself is, however, too faint to be detected (see Section
  \ref{sec:tail}.

Inside the central few AU, the effective temperature of the asteroid is larger
than the melting and evaporation temperature of iron ($T_{\rm Fe,m} \simeq
1800$ K, $T_{\rm Fe,v} \simeq 3100$ K) and the sublimation temperature of
carbon ($T_{\rm C,v} \sim 3900$ K; carbon does not have a liquid phase at
pressures below a few MPa). Therefore, the asteroid's outer layers should
indeed be evaporating as it is passing through the inner regions of the
accretion flow onto \sgra.

The radius at which temperature $T_{\rm X}$ is reached is given by
\begin{equation}
 \label{eq:rvap}
R_{\rm X} \simeq \left(\frac{T_{\rm X}}{1.4 \cdot 10^4
  \rho_{18}^{1/4}}\right)^{-8/(3+2s)} \; \mathrm{AU},
\end{equation}
where $T_{\rm X}$ is one of the sublimation temperatures of interest as above.
For the three cases of interest,  and $s = 1.23$ we find
\begin{equation}
R_{\rm Fe,m} \simeq 21 \; \mathrm{AU},\; R_{\rm Fe,v} \simeq 10 \;
\mathrm{AU},\; R_{\rm C,v} \simeq 7 \; \mathrm{AU.}
\end{equation}
This shows that asteroids start melting and evaporating at $R\sim 10$ AU,
i.e., well outside the tidal disruption radius. Of the two materials, we
expect carbon to be more abundant, so we use its parameters in subsequent
calculations.

To calculate the mass loss by the asteroid, we follow the classical
meteor ablation considerations (\citealt{Bronshten1983book}, see also \S
2.3.2 in \citealt{Alibert2005A&A}), which give
\begin{equation}
\label{mdot_vap1}
\dot{M}_{\rm v} \sim  \frac{\pi r^2 \Phi_{\rm a} C_{\rm H}}{2 Q_{\rm C,v}}.
\end{equation}
Here, $\dot{M}_{\rm v}$ is the mass loss rate due to vaporisation and
$Q_{\rm C,v} \sim 3.0 \cdot 10^{11}$ erg g$^{-1}$ is the energy per
unit mass required to raise the asteroid temperature to the
vaporisation temperature and evaporate it (the latter process is
energetically dominant). $C_{\rm H} < 1$ is an unknown dimensionless
coefficient which specifies how much of the bulk mechanical energy
inflow into the asteroid goes into the mass loss as opposed to thermal
re-radiation of that flux. In the high density environment of Earth and
Jupiter atmospheres, $C_H$ can be very small because the optical depth
of the evaporating material can be large and hence the asteroids
self-shield themselves  efficiently (so-called ``vapor
shielding''). For example, for asteroids of size $1-10$~m in the
Earth's atmosphere, $C_{\rm H} \sim 10^{-3}$ \citep{Svetsov1995Icar},
but this value increases with altitude (i.e. with decreasing
atmospheric density).

In the very low ambient gas density environment we study, $C_H$ is
likely to be close to unity because the optical depth of the self-shielding 
material is small for two reasons. Firstly, the evaporating
gas may be heated up to temperatures of the order of that of the
surrounding medium, which is $10^9-10^{11}$ K, at which point it would
be completely ionised and only electron scattering opacity would be
important. Secondly, the column depth of the evaporating flow is not
large. To see this, assume that evaporated gas outflows at $v_{\rm
  ev}\sim 10$ km/s. Since $\dot M_{\rm v} = 4\pi r^2 \rho_{\rm v} v_{\rm
  ev}$, the column depth of self-shielding material is
\begin{equation}
\Sigma_{\rm v} \sim \rho_{\rm s} r = \frac{\dot M_{\rm v}}{4\pi r v_{\rm ev}}\;.
\label{sigmav}
\end{equation}
Using equation \ref{mdot_vap1} we have,
\begin{equation}
\Sigma_v \sim \frac{C_H r \Phi_a}{2 Q_{\rm C,v} v_{\rm ev}} \approx
0.3 C_H \frac{r}{1 \hbox{km}}\; {\rm g cm}^{-2}.
\label{sigma_num}
\end{equation}
With opacity coefficient not too different from electron scattering, the
evaporated material is obviously optically thin. We hence conclude that
thermal ablation of asteroid fragments should be very effective with $C_H\sim
1$ for fragment size $r \simlt 1 $ km.

The evaporation rate is
\begin{equation}
\label{mdot_vap}
\dot{M}_{\rm v} \sim 1.3 \cdot 10^{13} \rho_{18} \; R_{\rm
  AU}^{-3/2-s} \; r_1^2\; C_H \; \mathrm{g \; s^{-1}}\;.
\end{equation}
For convenience, we define the vaporization timescale,
\begin{equation}
t_{\rm v} = \frac{M_{\rm a}}{\dot{M}_{\rm v}} \sim 3.2 \cdot 10^5
\rho_{18}^{-1} \; R_{\rm AU}^{3/2+s} \; r_1 \; C_H^{-1} \; \mathrm{s}.
\end{equation}
We see that the smaller the asteroid, the faster it vaporises. The
material ablated from the asteroid might assume a cometary shape, with
a long gaseous tail  behind the solid head (cf. \S
\ref{sec:tail}).

\subsection{Total and partial asteroid disruptions}\label{sec:total}

We can now delineate the parameter space for the possible outcomes of an
asteroid's flyby near a SMBH. An asteroid on a parabolic orbit around
\sgra\ with pericentre distance $R_{\rm p}$ spends a time
\begin{equation}
t_{\rm fly} \simeq \pi t_{\rm d} = \pi \sqrt{\frac{R_{\rm p}^3}{2 G M_{\rm
      BH}}} \simeq 5600 R_{\rm AU}^{3/2} \; \mathrm{s}
\label{tfly}
\end{equation}
at radial distance comparable with $R_{\rm p}$. The ratio between the
vaporisation timescale and the flyby time is 
\begin{equation}
\frac{t_{\rm v}}{t_{\rm fly}} \sim 57 \rho_{18}^{-1} \; R_{\rm AU}^s
\; r_1 \; C_H^{-1}.
\label{tratio}
\end{equation}
This ratio is important in determining what exactly happens to an
asteroid as it swings by the SMBH.


\subsubsection{Orbits outside 1 AU but inside $\sim$ 10 AU}\label{sec:outside1}

For asteroids on orbits with pericenter distances larger than $R_{\rm
  td} \sim 1 $ AU (equation \ref{r_td}), the asteroid is not tidally
disrupted. If the orbit passes within $R_{\rm X} \sim 10$ AU
(eq. \ref{eq:rvap}), the surface layers of the asteroid are vaporised
at the rate given by equation \ref{mdot_vap}. Only a fraction $\sim
t_{\rm fly}/t_{\rm v}$ of the asteroid is ablated during the close
passage. Therefore, large asteroids passing \sgra\ farther away than 1
AU remain relatively untouched and leave the SMBH vicinity on their
initial parabolic orbits.

The luminosity released by material lost by the asteroid in this
regime can be estimated as
\begin{equation}
 \label{eq:lfout} L_{\rm f,out} = \xi \dot{M}_{\rm v} c^2 = 2 \cdot
10^{33} \; \xi_1 \; \rho_{18} \; R_{\rm AU}^{-3/2-s} \; r_1^2 \; \mathrm{erg
  \; s}^{-1}\;.  
\end{equation} 
For an approach distance of $5$ AU, this luminosity becomes observable
(i.e. $L_{\rm f,out} > 10^{34}$~erg s$^{-1}$) only if the asteroid
radius is $r \gtrsim 190$~km. Such large asteroids are rare. Thus
asteroids passing \sgra\ at pericenter distances larger than $\sim 1$
AU are unlikely to result in observable flares.

\subsubsection{Total destruction of asteroids inside 1 AU}\label{sec:inside1}

Inside the tidal disruption radius, the asteroid breaks into fragments
with sizes smaller than $r_{\rm frag} \sim 1$~km (cf. \S
\ref{sec:td}). For these smaller asteroid fragments, vaporisation is
much more efficient. The incoming remnants heat up, melt and vaporise
rapidly. This leads to a decrease in the material tensile strength,
allowing further fragmentation due to tidal shear.  As a result, most
of the asteroid's mass evaporates during the flyby (cf. equation
\ref{tratio}).  We estimate the luminosity as
\begin{equation}
 \label{eq:lfin}
L_{\rm f,in} = \frac{\xi M_{\rm a} c^2}{t_{\rm fly}} = 6 \cdot 10^{34} \;
\xi_1 \; R_{\rm AU}^{-3/2} \; r_1^3 \; \mathrm{erg \; s}^{-1}.
\end{equation}
At an approach of $1$ AU, this luminosity becomes observable for
asteroids of radius $r \gtrsim 10$ km. 

We can also estimate the maximum flare luminosity. If the asteroid mass is
larger than the total mass of the gas in the accretion flow inside $1$ AU,
then efficiency of converting the asteroid's bulk motion into radiation must
be reduced. Even if the massive asteroid is vaporised completely, the mass of
the quiescent accretion flow is simply not high enough to stop the evaporated
material bodily. The latter would continue on its outward course from the
inner 1 AU. A part of the disrupted material comes back to \sgra\ as in the
stellar disruption case but with a time delay much longer than the dynamical
time in the inner AU. The proper estimate for the luminosity is then much
smaller than equation \ref{eq:lfin} suggests.

This sets an upper limit to the mass of an asteroid that is wholly
disrupted and stopped in the inner AU:
\begin{equation}
M_{\rm a,max} \lesssim M_{\rm g}\left(R < R_{\rm td}\right) \simeq 6.7 \cdot
10^{22} \; \rho_{18} \; \mathrm{g}\;,
\end{equation}
for $s = 1.23$, yielding radius $r \sim 250$ km. The luminosity
that an asteroid this massive would produce if it evaporated is
\begin{equation}
 \label{eq:lmax}
L_{\rm f,max} = \frac{\xi M_{\rm g}\left(R < R_{\rm td}\right) c^2}{t_{\rm
    fly}} \approx 10^{39} \; \xi_1 \; \rho_{18} \; R_{\rm AU}^{-3/2} \;
\mathrm{erg \; s}^{-1}.
\end{equation}
No flares of this magnitude have been detected so far, but this may be quite
reasonable as such large asteroids are expected to be rare.

\subsection{Summary on asteroid disruption}\label{sec:pod_summary}

From the arguments outlined above, we see that any large asteroids
passing \sgra\ within $R \sim 1$ AU qould be tidally disrupted
and efficiently vaporised. If their material is mixed with the
background accretion flow, the bulk kinetic energy of their orbital
motion around \sgra\ would be deposited into the accretion flow
around the SMBH. If the asteroid's initial radius exceeds $\sim
10$~km, this energy deposition might be large enough to produce an
observable flare.

Asteroids passing at larger $R$, on the other hand, are not tidally
disrupted. Their vaporisation times are longer than the time they spend near
the pericenters of their orbits. Therefore, they lose just a small fraction
of their mass. The amounts of mass and energy deposited by such more distant
flybys in the inner regions near \sgra\ are small, and thus no bright flares
from such passages could be produced.

\section{Flare frequency and luminosity distribution} \label{sec:supply}

\subsection{The ``Super-Oort cloud'' of asteroids}\label{sec:reservoir}

\citep{Nayakshin2011arXiv} have recently suggested that AGN may be
surrounded by several-pc scale clouds of asteroids and planets that have been
formed in situ. In this model, star formation episodes take place inside a
massive self-gravitating AGN accretion disc \citep{Paczynski1978AcA,
  Kolykhalov1980SvAL, Collin1999A&A, Goodman2003MNRAS, Paumard2006ApJ,
  Nayakshin2007MNRAS} during gas-rich phases when the super-massive black hole
grows rapidly. The AGN disc orientation performs a random walk due to chaotic
mass deposition events of individual large gas clouds \citep[as argued
  by][]{King2006MNRAS, Nayakshin2007arXiv, Hobbs2011MNRAS}. As a result, a
kinematically and geometrically thick cloud of stars surrounds the SMBH over
time. The asteroids are then stripped from their parent stars by close
passages of perturbers, such as other stars or stellar remnants, or by tidal
forces of the SMBH. This creates a geometrically thick torus of asteroids and
planets which may be called a ``Super-Oort cloud'' of SMBH by analogy with the
Oort cloud of the Solar System.

To estimate the properties of this cloud as relevant to our goals here, we
first consider asteroids at birth of a single star, assuming that their
population is not too dissimilar from that found in ``debris discs'' of nearby
stars and the Solar System. Physically, asteroids are remnants of
protoplanetary discs and the planet formation process in stellar
systems. While the planet formation process is itself not yet understood, we
may use observational constraints on the properties of debris discs around
nearby stars. Let $n(r)$ be the differential distribution function of
asteroids, so that the number of asteroids with radii between $r$ and $r+dr$
is
\begin{equation}\label{eq:dndr}
n\left(r\right) dr = n_0 \left( \frac{r}{r_0} \right)^q dr,
\end{equation}
where the slope $q$ can be reasonably expected to vary between $-3$ and $-4$,
but is probably close to the value $-3.5$ expected if the asteroid population
is the high-mass tail of a collisionally evolved debris disc
\citep{Wyatt2008ARA&A}. We now calibrate $n_0$ by requiring that the total
mass of asteroids per star is $M_{\rm a,t.}$:
\begin{equation}\label{eq:mkb}
M_{\rm a,t.} = \int_{r_{\rm min}}^{r_{\rm max}} M_{\rm a}(r) n(r) dr \simeq
\frac{4 \pi \rho_{\rm a}}{3 \left(q+4\right)} r_{\rm max}^{q+4}
\frac{n_0}{r_0^q},
\end{equation}
where we have assumed that $q > -4$, and therefore it is the upper limit of
the distribution that is more important. We now find the total number of
asteroids with radius $r > r_{\rm X}$ per star:
\begin{equation}\label{eq:fa}
\begin{split}
f_{\rm a}\left(r > r_{\rm X}\right) &= \int_{r_{\rm X}}^{\infty} n(r) dr =
\frac{n_0}{r_0^q} \frac{1}{-q-1} r_{\rm X}^{q+1} = \\ &= \frac{3 M_{\rm
    a,t.}}{4 \pi \rho_{\rm a}} \frac{q+4}{-q-1} \frac{r_{\rm X}^{q+1}}{r_{\rm
    max}^{q+4}}.
\end{split}
\end{equation}

The mass in asteroids/solid bodies per star, $M_{\rm a,t.}$, is not
easily constrained at present. First of all, the absolute upper limit
for this quantity is the total metal (dust) content of a protostellar
disc, which is of the order of $10^{-3}\msun$ \citep[assuming Solar
  metallicity and the disc mass of $\sim 0.1\msun$; see also a
  compilation of dust mass observations in Figure 3
  of][]{Wyatt2008ARA&A}. The minimum mass of the asteroid population,
on the other hand, is the mass of dust in debris disc systems. The
dust particles in these aged populations are rapidly blown away by the
radiation of the parent stars, and must be replenished by a credible
source. The collisional cascade that grinds asteroids into the
microscopic dust is believed to be such a source. Figure 3 of
\citet{Wyatt2008ARA&A} shows that the dust mass for observed debris
disc is of the order $\sim (10^{-8}-10^{-7})\msun$. The minimum mass
of the asteroids in these discs should be at least several orders of
magnitude higher.

Given this, we take the total mass of the asteroids per star as a free
parameter of the model, setting $M_{\rm a,t.} = 10^{-5} m_5 \msun$,
where $m_5$ is a dimensionless parameter which is hopefully not too
different from unity.  Setting $r_{\rm max} = 500$ km and $q = -3.5$
for illustrative purposes, we find
\begin{equation}\label{eq:fa2}
f_{\rm a}\left(r > r_{\rm X}\right) = 10^3 \; m_5 \left(\frac{r_{\rm X}}{500
  \mathrm{km}}\right)^{q+1} = 2 \cdot 10^{7} \; m_5 \; r_1^{q+1}.
\end{equation}
Thus there are approximately $2 \cdot 10^7$ asteroids {\it per star}
that may cause observable flares. Assuming the mean stellar mass
  inside the sphere of influence of \sgra\ is $\sim 1 \msun$ gives
  $N_* = 4 \cdot 10^6$ stars and a grand total of $N_{\rm a} \sim 8
  \cdot 10^{13} m_5$ asteroids large enough to cause observable
flares with the default parameter values chosen above.

\subsection{Event rates} \label{sec:rates}

\subsubsection{A quick estimate}\label{sec:rough}

Before proceeding to more detailed calculations, let us simply assume that the
spatial and velocity distribution of asteroids is exactly the same as that of
parent stars. As the mean density of a main sequence solar mass star is
similar to that of an asteroid, the tidal disruption radius for both is about
the same. Given that the expected rate of stellar tidal disruptions in the
Galactic Centre is $\dot{N_*} \sim 10^{-5}$ yr$^{-1}$, the rate for disruption
of asteroids is $\dot{N_*}$ times the number of asteroids ($r> 10$ km) per
star:
\begin{equation}\label{eq:dotncrude}
\frac{dN}{dt} \sim \dot{N_*} f_{\rm a} \sim 0.6 \; {\rm day}^{-1}
\left(\frac{\dot{N_*} m_5}{10^{-5}\; \hbox{yr}^{-1}}\right)\;.
\end{equation}
We see that we need $m_5 \simgt 1$ to satisfy the observed flare rates.

We can do an additional sanity check. If the currently observed
flaring rate is representative of a long-term quasi-static process,
then during the lifetime of the Galaxy, $t_{\rm Gal} \sim 10^{10}$ yr,
we expect $N_{\rm tot} \sim 3 \cdot 10^{12}$ flares to have
occured. This number is smaller than the total number of asteroids $r
> 10$ km as estimated above, $N_{\rm a} \sim 10^{14}$, within the
sphere of influence of \sgra.

\subsubsection{A filled loss cone estimate}

In order to make more detailed estimates of the asteroid disruption rates, we need
to calculate the evolution of the angular momentum distribution of the
asteroid population. In accordance with our simple model, given that
there are $f_a$ ``interestingly'' large asteroids per star, the number density
of asteroids inside \sgra\ sphere of influence is
\begin{equation}\label{eq:na}
n_{\rm ast} = n_* f_{\rm a},
\end{equation}
where $n_*$ is the number density of stars in the same region.

If the loss cone of the asteroid distribution in angular momentum and
energy space is kept full by some process, then the limiting rate of
events is given by the estimate of spherical collisionless
accretion. Following the derivation in Chapter 14.2 of
\citet{Shapiro1983bhwd.book} (their eqn. 14.2.19), the number
accretion rate onto a sphere of radius $R_{\rm t} = 1$ AU is
\begin{equation}
 \label{eq:madot}
\frac{dN}{dt} = \frac{2 \pi G M_{\rm BH} R_{\rm t} n_{\rm ast}}{\sigma},
\end{equation}
where $\sigma \simeq 10^7$ cm s$^{-1}$ is the velocity dispersion in
the Galactic bulge. Numerically,
\begin{equation}
n_{\rm ast} \simeq \frac{3 N_*}{4 \pi R_{\rm h}^3} f_{\rm a}
\simeq 7.6 \cdot 10^{-44} \; m_5 \; r_1^{q+1} \; \mathrm{cm}^{-3},
\end{equation}
where $R_{\rm h} \simeq 2$ pc is the radius of influence of \sgra. The number
accretion rate of asteroids onto \sgra\ is then
\begin{equation}
\begin{split}
\frac{dN (r > r_{\rm x})}{dt} &\sim 3.8 \cdot 10^{-4} \; R_{\rm AU} \; m_5 \;
r_1^{q+1} \; \mathrm{s}^{-1} \\ &= 33 \; R_{\rm AU} \; m_5 \; r_1^{q+1} \;
\mathrm{day}^{-1}.
\end{split}
\label{filled}
\end{equation}
This is a large rate which may not be realistic since it assumes a
filled loss cone.

\subsubsection{A depleted loss cone rate} \label{sec:losscone}

If the loss cone is almost empty, then the accretion rate is set by
its refilling timescale. The classical loss cone refilling arguments,
 e.g., \citet[][eqn. 6.11]{Alexander2005PhR} and references therein, give
\begin{equation}
 \label{eq:dndt}
\frac{dN}{dt} \sim \frac{2f_{\rm a}N_*}{{\rm ln}\left(R_{\rm h}/R_{\rm
    t}\right)t_{\rm r}(R_{\rm h})} \simeq 5 \cdot 10^{-12} f_{\rm a} \;
\mathrm{s}^{-1},
\end{equation}
where $t_{\rm r}(R_{\rm h}) \simeq 4 \cdot 10^9$ yr is the relaxation
time at $R_{\rm h}$.  Substituting for $f_{\rm a}$ from
eq. (\ref{eq:fa2}) gives
\begin{equation}
 \label{eq:dndt2}
\begin{split}
\frac{dN (r > r_{\rm x})}{dt} &\simeq 9.5 \cdot 10^{-5} m_5 \; r_1^{q+1}
\; \mathrm{s}^{-1} \\ &\simeq 8 \; m_5 \; r_1^{q+1} \;
\mathrm{day}^{-1}.
\end{split}
\end{equation}
This is somewhat smaller than estimate in equation \ref{filled}.

\subsection{Asteroid-asteroid collisions}\label{sec:collisions}

In the above treatment, we only considered gravitational perturbations of
asteroid orbits by stars (asteroids themselves are too small to perturb each
other's orbits gravitationally in the central parsec of the Galaxy). Asteroids
do collide bodily with each other, and some of these collisions can lead to
what is called a catastrophic collision (e.g., a collision which breaks the
asteroid into two or more pieces). Since we are interested in large bodies for
which fragmentation conditions depend on self-gravity rather than tensile
strength \citep{Wyatt2008ARA&A}, the size of an impactor that can just shatter
an asteroid of radius $r$ is derived from
\begin{equation} \frac{M_{\rm a}\left(r_{\rm
    i}\right) v_{\rm i}^2}{2} = \frac{G M_{\rm a}^2\left(r\right)}{r},
\end{equation}
where the subscript 'i' stands for 'impactor'. Expressing mass in
terms of asteroid radius gives an expression
\begin{equation}
 \label{eq:rimp}
r_{\rm i} \sim 1.9 \cdot 10^3 \; r_1^{5/3} \; v_{100}^{-2/3} \;
\mathrm{cm}, 
\end{equation} where the impactor velocity is
parametrised in units of $100$~km/s.  Now we consider a large asteroid
moving with velocity $v_{\rm i}$ through a stationary cloud of other
asteroids. By definition, it sees on average $1$ impactor large enough
to shatter it in a cylinder of area $\pi r^2$ and length $v_{\rm i}
t_{\rm coll}$, where $t_{\rm coll}$ is the collision timescale. Since
the number density of impactors can be expressed using
eq. (\ref{eq:na}), we have 
\begin{equation}
\begin{split}
t_{\rm coll} & = \left[f_{\rm a}(>r_{\rm i})n_* \pi r^2 v_{\rm i}\right]^{-1} = \\
&= 2.1 \cdot 10^9 \; m_5^{-1} \; r_1^{13/6} \; v_{100}^{-8/3} \; \mathrm{yr}.
\end{split}
\end{equation}
This timescale is longer than the Hubble time for $r > 24$
km. Therefore we see that while some of the smaller asteroids may be
destroyed, the largest ones, which also contain the majority of the
total mass, are not. Furthermore, the estimate assumes a steady-state
collisional fragmentation cascade of the form (\ref{eq:dndr}), which
may actually turn over at small $r$ if the smaller bodies are
removed from the cascade rapidly.

\subsection{Flare luminosity distribution} \label{sec:size}

The asteroid number density (eq. \ref{eq:fa2}) may be used to calculate the
number of asteroids per star that have mass greater than $M_{\rm X}$:
\begin{equation}
\label{eq:fam}
f_{\rm a}\left(M_{\rm a} > M_{\rm X}\right) = 6 \cdot 10^{9}\; m_5
\left(\frac{M_{\rm X}}{4 \cdot 10^{15} \mathrm{g}}\right)^{(q+1)/3}\;.
\end{equation}

The observed distribution of flare luminosities follows a $LN_{\rm L} \propto
L^\alpha$ law, with $-1 \lesssim \alpha \lesssim 0$ \citep[; see also \S
  \ref{sec:model}]{Dodds-Eden2011ApJ}. Using this, the frequency of flares
with luminosity $L_{\rm f} > L_{\rm X}$ is
\begin{equation}
N(L_{\rm f} > L_{\rm X}) = \int_{L_{\rm X}}^\infty N_{\rm L}dL \propto LN_{\rm
  L} \propto L^\alpha.
\end{equation}
Since the luminosity of a flare from an asteroid of mass $M_{\rm X}$ is
proportional to $M_{\rm X}$ in our model, we can convert the asteroid mass
distribution into flare lumunosity distribution:
\begin{equation} 
N(L_{\rm f} > L_{\rm X}) \propto L_{\rm X}^{(q+1)/3}, 
\end{equation} 
where the value of the exponent varies between $-2/3$ (for $q = -3$) and $-1$
(for $q = -4$). This is within the observationally
constrained range of $\alpha$ \citep{Dodds-Eden2010ApJ}.

 Flares with luminosity $L_{\rm f,X} =
10^{34} \; L_{34}$ erg s$^{-1}$ correspond to
\begin{equation}
r \sim 10 \; \xi_1^{-1/3} \; R_{\rm AU}^{1/2} \; L_{34}^{1/3} \mathrm{km}.
\end{equation}
Using eq. (\ref{eq:dndt2}), we normalize the flare luminosity distribution, and obtain, for $q= -3.5$ as the likely value, 
\begin{equation}
\dot{N} \sim 8 \; m_5 \; L_{34}^{-5/6} \mathrm{day}^{-1}.
\label{dndl}
\end{equation}
The brightest flare seen so far has $L_{\rm X, max} \sim 10^{36}$ erg
s$^{-1}$, requiring $r \gtrsim 45$ km, which corresponds to $\dot{N} \sim 0.2$
day$^{-1}$. The total duration of {\it Chandra} observations of \sgra\ is
$t_{\rm obs} \sim 1.4$ Msec, so we expect it to have seen $N \lesssim 3.5$
flares of this magnitude or brighter, which is not too far off from the
one flare per day actually observed.

\section{Planet disruptions}\label{sec:planets}

Although much less frequent, planet disruptions may also occur near
\sgra. Their frequency is probably comparable to that of stellar disruptions,
e.g., one per $\sim 10^5$ yrs \citep{Alexander2005PhR}, if we assume one
planet per star on average. Consider now a gas giant planet passing within 1
AU of \sgra. Its disruption is quite analogous to that of a star. The most
bound disrupted material is on an orbit with a semimajor axis
(cf. eq. \ref{eq:semi-axis})
\begin{equation} \label{eq:semi-axis-planet}
a_{\rm p} \sim R \frac{v_{\rm a}}{v_{\rm esc,p}}\; \sim 2 \cdot 10^3
\;{\rm AU} \sim 0.01 \; {\rm pc},
\end{equation}
where $v_{\rm esc,p}$ is the escape velocity from the planet's surface 
($\sim 60$~km/s for a Jupiter mass body). The bound debris returns back to
the vicinity of \sgra\ after a time
\begin{equation}
P_{\rm orb} \sim 2 \pi \sqrt{\frac{a_{\rm p}^3}{G M_{\rm BH}}} \sim 30 \; {\rm yr}.
\end{equation}
The maximum fallback rate is thus $\dot M_{\rm back} \sim 10^{-3} \msun
/(30$yr)~$=3\times 10^{-5} \msun$~yr$^{-1}$. This rate is significantly larger
than the estimated current quiescent accretion rate onto \sgra, $\dot M \sim
10^{-8} \msun$~yr$^{-1}$. Conceivably one could expect \sgra\ to brighten by
multiple orders of magnitude for $\sim$ tens to a hundred years. The maximum
bolometric luminosity is obtained assuming the radiatively efficient
conversion of accretion energy into radiation:
\begin{equation}
L_{\rm back} \le 0.1 c^2 \dot M_{\rm back} \sim 2 \times 10^{41} \; {\rm erg
  s}^{-1}\;.
\label{lback}
\end{equation}

The order of magnitude of this luminosity and the flare duration (tens
of years) are within that inferred to have occured some $\sim 300$ yrs
ago, when \sgra\ was apparently as bright as $\simgt 10^{39}$ erg
s$^{-1}$ in X-rays \citep{Revnivtsev2004A&A,Terrier2010ApJ}. We
speculate that tidal disruption of a rogue gas giant planet could
account for that activity episode.

\section{Emission mechanisms} \label{sec:flare}

A detailed modeling of the emission from the vaporised material mixed with the
background flow is beyond the scope of our paper due to many physical
uncertainties (such as the role of magnetic fields along the interface between
the vaporised tail and the ambient gas). However, it is possible to rule out
several potential emission mechanisms and point out the most promising
scenario under which tidal disruption of asteroids could produce the
spectra consistent with those observed.

\subsection{Asteroid disruptions are not "accretion rate" flares}

The simplest view on emission from asteroids is that they bring in an
additional mass to the inner accretion flow onto \sgra. The transient
enhancement in the accretion rate onto \sgra could then make it temporarily
brighter. However, in \S \ref{sec:inside1}, we pointed out that the mass of
the background quiescent accretion flow onto \sgra\ inside 1 AU is $\sim
10^{23}$ g based on the model of \cite{Yuan2003ApJ}. This is $\sim 3$ orders
of magnitude heavier than the typical asteroid mass that we considered here
(see \S \ref{sec:inside1}). The mass added by an asteroid to the region within
1 AU is simply too small to make an accretion powered flare unless the
asteroid's diameter is about 500 km, which must be a very rare
event. Therefore, if asteroid tidal disruptions are to be observable, they are
to be accompanied by production of particles emitting differently (more
efficiently) than the backround radiatively inefficient accretion flow.

This is consistent with observational constraints on the flares.
\citet{Markoff2001A&A} have shown that constraints on the absence of
significant variability in the radio emission of \sgra\ suggest that during
the flares it is not the magnetic field but rather the energy distribution of
emitting particles that vary. This conclusion rules out accretion-powered
flares as the mean magnetic field is expected to be proportional to the flow
pressure and thus density. Similarly, \citet{Yuan2004ApJ} found that infrared
flares from \sgra\ are best explained by assuming that a small fraction of
electrons in the flow (e.g., a few percent) is accelerated into a non-thermal
power-law tail.

\subsection{Thermal radiation from the asteroid's tail}\label{sec:tail} 

One new population of particles, compared with the very hot $T\sim 10^{11}$ K
quiescent accretion flow, is in the the vaporising asteroid's ejecta while it
is still relatively cold, i.e., $T\sim T_{\rm X} \sim 10^4$~K (cf. \S
\ref{sec:vapor}). The ejecta has initially a much higher density than the
ambient medium and must expand into the latter as it heats up. The evaporating
coma is probably shaped as a conical tail behind the asteroid. Since the
surface area of the tail is much larger than the asteroid itself, the tail
should be much brighter than the asteroid's face, and perhaps observable from
Earth. Its emission can be approximated as thermal, since the thermalisation
timescale of electrons in the coma is \citep[e.g.,][]{Stepney1983MNRAS} less
than $1$~s. The bolometric luminosity of the emission emanating from the tail
is
\begin{equation}
 L_{\rm bb} \simeq A \; \sigma_{\rm SB} \; T_{\rm tail}^4 \; \frac{\tau}{\tau +1}, 
\end{equation}
where $\tau$ is the optical depth in the direction perpendicular to the tail
and $A$ is the surface area of the tail. The optical depth, $\tau$, is
\begin{equation}
\tau \sim \kappa \; \rho_{\rm tail}\; r_{\rm tail},
\end{equation}
with $\kappa \equiv 10 \kappa_1$ the opacity of the material. Assuming $A \sim
\pi r_{\rm tail} h_{\rm tail}$, where $r_{\rm tail}$ and $h_{\rm tail} \gg
r_{\rm tail}$ are the base radius and height of the cone, we note that
\begin{equation}
A \tau = \kappa \;\rho_{\rm tail} \;\pi r_{\rm tail}^2 h_{\rm tail} \sim \kappa \;
M_{\rm tail},
\end{equation}
where $M_{\rm tail} \lesssim M_{\rm a}$ is the mass of the tail. The maximum
thermal luminosity from the evaporating ejecta is achieved if the tail is
moderately optically thin, $\tau \simlt 1$, and is
\begin{equation}
L_{\rm bb,max} \simeq \kappa \; M_{\rm a} \; \sigma_{\rm SB} \; T_{\rm tail}^4 \simeq 2
\cdot 10^{31} \;\kappa_1 \;r_1^3 \;\left(\frac{T_{\rm tail}}{10^4\;{\rm
    K}}\right)^4 \;{\rm erg \;s}^{-1}.
\end{equation}
This value is significantly smaller than the quiescent NIR luminosity
of \sgra. Further, the blackbody spectrum for $T_{\rm tail} \sim 10^4$
K peaks in the UV, where extinction is very large. In the NIR
frequencies, where \sgra\ line-of-sight is less obscured, the tail
emits in the Rayleigh--Jeans regime and hence is far dimmer than the
bolometric luminosity estimate above. Summing this up, we conclude
that direct thermal radiation from the evaporating material is not
observable against the background of quiescent \sgra\ emission.

On the other hand, the opacity of the expanding tail may be
sufficiently large to account for absorption of the quiescent
emission, producing the occasionally observed dimming of \sgra\ in
radio and sub-mm wavelengths just before a flare
\citep{Yusef-Zadeh2010ApJ}.

\subsection{A new relativistic population of particles?}

The particles in the asteroid's tail do have a very different velocity
distribution compared with that of the background flow.  The initial
velocities of the ions in the tail are strongly dominated by the bulk motion
inhereted from the initial's asteroid's orbit around \sgra.  This velocity is
somewhat larger than the ion sound speed of the accretion flow (for a hot
quasi-spherical inflow the sound speed is of the order of the local Keplerian
speed).

When the vaporised tail particles get mixed with the accretion flow particles,
we get a very anisotropic velocity distribution.  Therefore we expect a number
of plasma instabilities to operate while the ions and the electrons of the
vaporised material are assimilated into the hot \sgra\ accretion flow. If
non-thermal electrons reach equipartition with the shocked ions as in the
models of gamma-ray bursts \citep[e.g.,][]{Meszaros1994ApJ}, their maximum
$\gamma$-factors could be as large as $(G\mbh/c^2R)\, (m_i/m_e) \sim 0.1
m_i/m_e$, where $m_i$ and $m_e$ are the ion and electron mass
respectively. Even for $m_i=m_p$, inside the inner AU this factor exceeds
100.

It is thus likely that a disrupted asteroid produces a transient population of
high energy electrons along its original trail. This population should cool by
radiative emission and mixing with the background. Without going into a
model-dependent characterisation of these processes, we only note that a tidal
disruption event may plausibly give rise to the hot particle distributions
needed in the scenarios of transient plasma blob based flare emission
\citep[e.g. ][]{Trap2011A&A}.  We feel that this scenario of converting
  asteroid's bulk energy into radiation is by far the most promising one to
produce spectra resembling \sgra\ flares.

\section{Discussion and conclusions} \label{sec:discuss}

In this paper we considered the fate of asteroids passing \sgra\ within a few
AU on nearly radial orbits.  As noted in the Introduction, we are unable to
make detailed spectral predictions at this time, but we do obtain interesting
constraints on the energetics, bolometric luminosity and frequency of flares
powered by tidal disruption of asteroids. We give a short summary of our
results and model predictions here.

The physical picture of an asteroid disruption near \sgra\ has two
stages. Firstly, the asteroid is tidally disrupted if it enters the inner
$\sim 1$ AU region, where it is broken into smaller fragments bound by
molecular forces rather than gravity. These fragments are probably less than a
few hundred meters in radius. The second stage of the disruption is
evaporation of these smaller fragments by heat released due to aerodynamic
friction of the fragments on the quiescent accretion flow new \sgra. The bulk
kinetic energy of the asteroid is sufficient to power an observable flare if
the asteroid's radius is greater than about 10~km.  We then estimated the
asteroid disruption events rate based on the assumption that the number of
asteroids per star is reasonably large and is of the order of that inferred
from nearby stars.

Our model makes the following predictions:

1. The small size of the flaring region, $R_{\rm f} \lesssim 10 R_{\rm S} \sim
1$~AU. This is the tidal disruption radius for a typical asteroid. Bodies
passing \sgra\ outside this radius lose some mass by vaporisation of the
outer layers, but the amount of such a mass loss is too small to give a
detectable flare (cf. \S \ref{sec:pod_summary}).

2. Frequency of flare occurence is given by the rate at which
asteroids from the "Super-Oort" cloud in the inner parsec
\citep{Nayakshin2011arXiv} are deflected onto low angular momentum
orbits that bring them within the tidal disruption radius. For
fiducial numbers, our model yields a reasonable agreement with the
observations (\S \ref{sec:rates}). This estimate however sensitively
depends on the poorly constrained normalisation factor $m_5$ (equation
\ref{eq:dndt2}).

3. The model naturally predicts a wide range of flare luminosities due to a
range in asteroid sizes. Under the assumption that flare luminosity is
proportional to the mass of the asteroid disrupted, we also find that the
luminosity-frequency relation for flares is within observational constraints
(\S \ref{sec:size}).

4. Extending the model to tidal disruption of gas giant planets predicts rare
but much brighter flares. One such event may have produced the AGN-like flare
of \sgra\ $\sim 300$ years ago (\S \ref{sec:planets}).

5. The flare frequency in our model is given by the supply of asteroids rather
than by the properties of the hot quiescent flow. Therefore, we would expect
no strong correlation between the quiescent properties of \sgra\ spectrum and
the occurence of flares (that is, if \sgra\ quiescent emission were to
brighten or dim by a factor of a few in the next few years, we would not
expect the rate of flaring to be affected). A weak correlation may be expected
if the luminosity -- asteroid mass relation is not quite linear as assumed
here.

6. We also note that asteroid disruption flares from exceptionally large
asteroids may be observable from nearby galactic nuclei. Equation \ref{dndl}
predicts that a flare with $L\sim 10^{39}$ erg s$^{-1}$ would occur every few
years at best. However, for a large enough sample of sources such events may
be detected in dormant nearby galactic nuclei.

7. The external origin of the flare trigger provides a way to test this
model. \citet{Markoff2005ApJ} showed that in the
flaring state, \sgra\ sits on the Fundamental Plane of radio and X-ray
luminosities for black holes (both stellar mass and supermassive). The
Fundamental Plane is thought to arise due to accretion physics, so if the
flares are caused by accretion instabilities of any kind, flares more luminous
than $L_{\rm X} \simeq 10^{36}$~erg s$^{-1}$ should be accompanied by a
corresponding increase in radio luminosity, with a possible lag of months to
years. On the other hand, asteroid-induced flares should not exhibit this
correlation, at least not up to luminosities $L_{\rm f,max} \simeq
10^{39}$~erg s$^{-1}$ (eq. \ref{eq:lmax}), when the asteroid mass becomes
comparable to the gas mass in the quiescent flow. Future long-duration
  observational campaigns of \sgra\ may thus help distinguish between
  differing flare scenarios.

The least constrained parts of the model have to do with the exact
distribution of asteroids and their orbits in the hypothesised
"Super-Oort cloud" around \sgra, and with conversion of the bulk
kinetic energy of the asteroids into electromagnetic radiation. 
  However, there almost certainly are asteroids in the central few pc
  of the Galaxy and the processes described here must occur. Our paper
  makes several estimates of the effects that asteroids have on the
  luminosity of \sgra\ and suggests a method to distinguish between
  such externally caused flares and accretion-instability caused
  ones. If future observations reveal that asteroid disruptions are
  responsible for at least a fraction of the flares, this would be an
  important step in understanding the accretion processes in \sgra. In
  addition, further investigation may help constrain the size of the
  asteroid population in the Galactic centre.

\section*{Aknowledgment}

Theoretical astrophysics research at the University of Leicester is
supported by a STFC rolling grant. KZ acknowledges a STFC
studentship. SM is grateful for support from a Netherlands
Organization for Scientific Research (NWO) Vidi Fellowship, as well as
The European Community's 7th Framework Programme (FP7/2007-2013) under
grant agreement number ITN 215212 Black Hole Universe.

\bibliographystyle{mnras}
\bibliography{asteroids-ref}

\end{document}